\documentclass[useAMS,usenatbib]{mn2e}

\usepackage{graphicx}
\usepackage{color}
\usepackage{adjustbox}

\newbox\grsign \setbox\grsign=\hbox{$>$} \newdimen\grdimen \grdimen=\ht\grsign
\newbox\simlessbox \newbox\simgreatbox
\setbox\simgreatbox=\hbox{\raise.5ex\hbox{$>$}\llap
     {\lower.5ex\hbox{$\sim$}}}\ht1=\grdimen\dp1=0pt
\setbox\simlessbox=\hbox{\raise.5ex\hbox{$<$}\llap
     {\lower.5ex\hbox{$\sim$}}}\ht2=\grdimen\dp2=0pt

\def\simless{\mathrel{\copy\simlessbox}}
\newbox\simppropto
\setbox\simppropto=\hbox{\raise.5ex\hbox{$\sim$}\llap
     {\lower.5ex\hbox{$\propto$}}}\ht2=\grdimen\dp2=0pt

\title[Chemical Abundances of GCs]{APOGEE Chemical Abundances of 
Globular Cluster Giants in the Inner Galaxy
} 
\author[Schiavon et al.]
{Ricardo P. Schiavon$^{1}$\thanks{E-mail: R.P.Schiavon@ljmu.ac.uk},
Jennifer A. Johnson$^{2}$,
Peter M. Frinchaboy$^{3}$,
\newauthor
Gail Zasowski$^{4}$,
Szabolcs M\'esz\'aros$^{5}$,
D.~A. Garc\'\i a-Hern\'andez$^{6,7}$,
Roger E. Cohen$^{8}$,
\newauthor
Baitian Tang$^{8}$,
Sandro Villanova$^{8}$,
Douglas Geisler$^{8}$,
Timothy C. Beers$^{9}$,
\newauthor
J.~G. Fern\'andez-Trincado$^{10}$,
Ana E. Garc\'\i a P\'erez$^{11}$,
Sara Lucatello$^{12}$,
\newauthor
Steven R. Majewski$^{11}$,
Sarah L. Martell$^{13}$,
Robert W. O'Connell$^{11}$,
\newauthor
Carlos Allende Prieto$^{6,7}$,
Dmitry Bizyaev$^{14,15}$,
Ricardo Carrera$^{6,7}$,
\newauthor
Richard R. Lane$^{16}$,
Elena Malanushenko $^{14}$,
Viktor Malanushenko $^{14}$,
\newauthor
Ricardo R. Mu\~noz$^{17}$,
Christian Nitschelm$^{18}$,
Daniel Oravetz$^{14}$, 
Kaike Pan$^{14}$,    
\newauthor
Alexandre Roman-Lopes$^{19}$,
Matthias Schultheis$^{20}$
\& Audrey Simmons$^{14}$.    
\\
$^{1}$Astrophysics Research Institute, Liverpool John Moores University,
     146 Brownlow Hill, Liverpool, L3 5RF, United Kingdom \\
$^{2}$Department of Astronomy, The Ohio State University, Columbus,
	OH 43210, USA \\
$^{3}$Texas Christian University, Fort Worth, TX 76129, USA \\
$^{4}$Department of Physics and Astronomy, Johns Hopkins University, 
    Baltimore, MD 21218, USA \\
$^{5}$ELTE Gothard Astrophysical Observatory, H-9704 Szombathely, Szent
     Imre Herceg st. 112, Hungary\\
$^{6}$Instituto de Astrof\'\i sica de Canarias, E-38205 La Laguna, Tenerife,
     Spain\\
$^{7}$Departamento de Astrof\'\i sica, Universidad de La Laguna (ULL),
     E-38206 La Laguna, Tenerife, Spain\\
$^{8}$Departamento de Astronom\'\i a, Casilla 160-C, Universidad de
     Concepci\'on, Concepci\'on, Chile \\
$^{9}$Dept. of Physics and JINA Center for the Origin of the Elements, 
     University of Notre Dame Notre Dame, IN  46530  USA\\
$^{10}$Institut Utinam, CNRS UMR6213, Universit\'e de Franche-Comt\'e, OSU
     THETA Franche-Comt\'e-Bourgogne, \\
     ~~Observatoire de Besan\c con, BP 1615, 25010 Besan\c con Cedex, France\\
$^{11}$Dept. of Astronomy, University of Virginia, Charlottesville,
	VA 22904-4325, USA \\
$^{12}$INAF-Osservatorio Astronomico di Padova, Vicolo dell’Osservatorio 5,
	I-35122 Padova, Italy \\
$^{13}$School of Physics, University of New South Wales, Sydney, NSW 2052,
     Australia\\
$^{14}$Apache Point Observatory and New Mexico State University, P.O. Box 59, \\
     Sunspot, NM 88349-0059, USA \\
$^{15}$Sternberg Astronomical Institute, Moscow State University, Moscow \\
$^{16}$Instituto de Astrof\'\i sica, Pontificia Universidad Cat\'olica de Chile, Av.
     Vicuna Mackenna 4860, 782-0436 Macul, Santiago, Chile \\
$^{17}$ Departamento de Astronom\'\i a, Universidad de Chile, Camino El 
     Observatorio 1515, Las Condes, Santiago, Chile \\
$^{18}$Unidad de Astronom\'\i a, Universidad de Antofagasta, Avenida Angamos 
     601, Antofagasta 1270300, Chile \\
$^{19}$Departamento de F\'\i sica, Facultad de Ciencias, Universidad de La Serena, 
     Cisternas 1200, La Serena, Chile \\
$^{20}$Laboratoire Lagrange (UMR7293), Universite de Nice Sophia Antipolis,
    CNRS, Observatoire de la C ´ ote d’Azur, \\
    ~~BP 4229, F-06304 Nice Cedex 4, France\\
}

\begin{document}

\date{Draft, 4 November, 2016}
\pagerange{\pageref{firstpage}--\pageref{lastpage}} \pubyear{2016}
\maketitle
\label{firstpage}
\begin{abstract}
We report chemical abundances obtained by SDSS-III/APOGEE for giant
stars in five globular clusters located within 2.2~kpc of the
Galactic centre.  We detect the presence of multiple stellar
populations in four of those clusters (NGC~6553, NGC~6528, Terzan~5,
and Palomar~6) and find strong evidence for their presence in
NGC~6522.  All clusters present a significant spread in the abundances
of N, C, Na, and Al, with the usual correlations and anti-correlations
between various abundances seen in other globular clusters.  Our
results provide important quantitative constraints on theoretical
models for self-enrichment of globular clusters, by testing their
predictions for the dependence of yields of elements such as Na,
N, C, and Al on metallicity.  They also confirm that, under the
assumption that field N-rich stars originate from globular cluster
destruction, they can be used as tracers of their parental systems
in the high-metallicity regime.
\end{abstract}

\begin{keywords}

\end{keywords}

\section{Introduction} \label{intro}

The discovery, within the past decade, of the presence of multiple
stellar populations in Galactic globular clusters (GCs) has forced
a revision of the traditional paradigm for the origin of these
objects.  For several decades, GC stars have been known to exhibit
Na, O, Al, and Mg \citep[e.g.,][]{gratton04,carretta10} spreads.
More recently, star-to-star variations in chemical-composition were
associated with the detection of multiple sequences in high-precision
colour-magnitude diagrams for the majority of Galactic GCs
\citep[e.g.,][]{piotto08}, leading to the suggestion of a complex
history of star formation and chemical enrichment in systems once
thought to be prototypical single stellar populations.  Under most
such scenarios, the so-called ``first-generation'' (FG) stars exhibit
abundance patterns that are similar to those of field stars of the
same [Fe/H], whereas the chemical compositions of ``second-generation''
(SG) stars depart from those patterns, showing enhancement in light
elements such as He, N, Na, and Al, and depletion in C, O, and
sometimes Mg.

No theoretical models based on the premise that GCs evolve chemically
have thus far been able to account for the extant data in detail
\citep[see][for a review]{renzini15}, so other alternatives have
been sought \citep[e.g.,][]{bastian13,hopkins14}, and those have
also been shown to fail \citep{bastian15}.  Naively, self-enrichment
models seem reasonable, for they frame GCs as low-mass manifestations
of processes of star formation and chemical enrichment known to
operate in galaxies---a notion that is particularly supported by the
detection of mass-chemical composition relations in GCs
\citep{carretta10,schiavon13,sakari16}.  However, to be tenable
they must rely on requirements that are not borne out by the data.
Chief amongst those is the so called ``mass budget problem'',
according to which, for any assumed initial mass function, the
observed mass of freshly produced nucleosynthetic material currently
observed in the atmospheres of SG stars requires a much larger
number of FG polluters than can be reconciled with the numbers of
existing low-mass FG cluster stars \citep[e.g.,][]{renzini08}---typically
by one or more orders of magnitude.

This assumption has been recently challenged \citep{schiavon16} by
the discovery of a large population of stars with enhanced N
abundances (N-rich stars) in the field of the inner Galaxy.  The
abundance patterns of these stars resemble those of SG stars in
Galactic GCs, suggesting that they are the possible leftovers of a
large population of early GCs that were entirely destroyed.
Interestingly, the maximum ratio between FG and SG stars in the
inner Galaxy is lower than required by models to solve the mass
budget problem \citep[see also][]{larsen14}.  Moreover, the lower
limit to the mass contained in stars originated in the presumptive
destroyed GCs is larger than that of the existing Galactic GC system
by an order of magnitude.  Since the metallicity distribution
function of N-rich stars is substantially different from that of
the present GC system \citep{harris96}, one is led to conclude that
the remaining GCs are not simply a scaled-down version of a much
larger precursor GC system.  Rather, the mass loss by the
existing GCs was probably modest, and the GC-like stars found in
the field today come from a parental GC population that was mostly
destroyed.  The latter conclusion, though potentially far reaching,
depends crucially on the assumption that N-rich stars are reliable
tracers of GC populations at all metallicities---in other words,
that SG stars in metal-rich GCs present the same levels of
enrichment/depletion in light elements as their more metal-poor
counterparts.

Progress in this field depends crucially on the mapping of the
multiple-population phenomenon across the entire volume of parameter
space covered by GCs, with metallicity being a particularly important
parameter.  Painstaking observational efforts have yielded a large
collection of colour-magnitude diagrams \citep[e.g.,][]{piotto02,piotto15}
and detailed abundance patterns of GC members
\citep[e.g.,][]{carretta10,pancino10,meszaros15}.  On the metal-rich
end, however, elemental abundances are not available for large
samples of GC members, because most metal-rich GCs are located in
the inner Galaxy, which is difficult to access in the optical due
to large dust extinction.  To our knowledge, only a very small
number of metal-rich GCs have been studied with sufficiently large
member samples to enable the detection of multiple stellar populations,
although low-resolution spectroscopy of stellar members
\citep[e.g.,][]{martell09,pancino10} and integrated spectroscopy
of extragalactic GCs \citep{schiavon13} suggests that they are
present.

It is important to document the presence of multiple populations
in metal-rich GCs for additional reasons.  Firstly, multiple
populations provide critical tests of stellar-evolution model
predictions for yields of light elements
\citep[e.g.,][]{karakas10,ventura13,dicriscienzo16} in a regime
that is important not only for the debate over globular cluster
formation, but also in the context of models for the chemical
evolution of galaxies \citep[e.g.,][]{pipino09}.  Secondly, an
estimate of the total mass contained in the presumptive dissolved
GCs discovered by \cite{schiavon16} in the inner Galaxy depends on
knowledge of the ratios between FG and SG stars across the entire
metallicity range.  Thirdly, there is evidence for the existence
of a positive correlation between the amplitude of abundance spreads
in GCs and their masses and metallicities, resulting from analysis
of the abundances of Na and O in individual Galactic GC stars
\citep{carretta10} and, in an indirect way, from the mean N abundances
of M31 GCs \citep{schiavon13}.  One would thus expect to find similar
abundance spreads in metal-rich Galactic GCs of moderate-to-high
masses.

High-resolution near-infrared (NIR) spectroscopy for large samples
of stars is probably the most efficient way to attack this problem.
The {\it Apache Point Observatory Galactic Evolution Experiment}
\citep[APOGEE,][]{majewski16} contributes importantly in that
regard.  A massive survey of Galactic stellar populations, APOGEE
obtained H-band $R \sim 22,500$ resolution spectra for over 150,000
stars, many of which are members of globular clusters.  This article
reports the abundance patterns of 23 candidate members of
five GCs situated within the inner Galaxy.  
Sample and data are described in Section~\ref{data}, and results
are presented in Section~\ref{results}.  Our conclusions are
summarised in Section~\ref{conclusions}.  A more exhaustive evaluation
of membership and an analysis of the detailed abundance patterns
of members of one of the GCs in our sample (NGC~6553) are presented
in a separate paper \citep{tang16}.  For a detailed analysis of
APOGEE abundances for members of GCs outside the inner Galaxy, we
refer the reader to \cite{meszaros15}.

\section{Data and sample} \label{data}

\subsection{Target Selection} \label{target}

The rationale behind targeting GC stars in APOGEE was twofold: on
the one hand, stars that were previously subjected to detailed
abundance analysis are useful for reality checks and potentially
the calibration of APOGEE elemental abundances.  On the other hand,
there is an obvious interest in expanding the database of GC elemental
abundances by targeting stars that are known or probable GC members,
but whose chemical compositions are unknown.  The range of
metallicities of our sample GCs overlaps largely with samples from
previous APOGEE studies \citep{meszaros15,schiavon16}, but does
include some of the most metal-rich GCs known in the Galaxy
(NGC~6528 and 6553).  Also targeted is a GC known to host at
least three populations with distinct [Fe/H] (Terzan~5), and two
more metal-poor GCs (NGC~6522 and Palomar~6).  The main relevant
properties of the target GCs are summarised in Table~\ref{properties}.
For each GC, lists of candidate members were put together including
targets from  both categories above, and these lists were fed to a
prioritization algorithm that assigned fibers to various targets
in each APOGEE field.  Fiber collision poses a major limitation
against a dense sampling of GC stars, by preventing the simultaneous
observation of targets separated by less than 1\arcmin.  However,
more stars could be observed from GCs located in fields that were
visited multiple times.  As a result of these constraints, the
sampling of the target GCs discussed in this paper is somewhat
serendipitous and not evenly distributed, ranging between one
candidate member for NGC~6522 and twelve for NGC~6553.  All targets
are giant stars that are likely GC members with stellar parameters
within the following range: $3600 \simless T_{\rm eff} \simless
4700~K$ and $0 \simless \log g \simless 2.6$.  For further details
on APOGEE target selection, we refer the reader to \cite{zasowski13}.

\begin{table*} \label{properties}
 \centering
\caption{Properties of the globular clusters targeted in this study.  Mean
radial velocities and tidal radii are taken from the 2010 edition of the
Harris (1996) catalogue.  For references on [Fe/H] and mass see text.
}
  \begin{adjustbox}{max width=\textwidth}
  \begin{tabular}{ccrrr}
\hline
\hline
ID   &  [Fe/H]  &  $<$RV$>$ (km/s) &  $r_t$ (arcmin) & Mass (M$_\odot$) \\
\hline
\hline
Palomar 6  &  --0.91  &  +181.0  &  8.3  &  $2.3 \times 10^5$ \\
Terzan 5  &  [--1.2,+0.3]  &  --82.0  &  6.7  &  $2.0 \times 10^6$ \\
NGC 6522  &  --1.00  &  --21.1 &  15.8  &  $6.0 \times 10^4$ \\
NGC 6528  &  --0.2   &  +206.6 &  4.1  &  $2.0 \times 10^5$ \\
NGC 6553  &  --0.2  &  --3.2 &  7.7  &  $3.0 \times 10^5$ \\
\hline
\end{tabular}
\end{adjustbox}
\end{table*}

\subsection{Data}

The results presented in this paper are based on the products of
Data Release 12 \citep[DR12,][]{alam15,holtzman15} of the SDSS-III/APOGEE
survey \citep{eisenstein11,majewski16}, consisting of accurate
elemental abundances and radial velocities, supplemented here by
2MASS astrometry \citep{skrutskie06} and GC structural parameters
and radial velocities from the 2010 edition of the Harris catalog
of Galactic globular clusters \citep{harris96,harris10}.  Elemental
abundances are based on the automatic analysis of APOGEE spectra
performed by the APOGEE Stellar Parameter and Chemical Abundances
Pipeline \citep[ASPCAP,][]{garcia16}, which performs a quantitative
comparison of observed spectra with a huge spectral library,
calculated on the basis of state-of-the-art model atmospheres
\citep{meszaros12,zamora15} and a comprehensive and accurate line
list \citep{shetrone15}.  The stellar spectra were themselves
collected with the APOGEE spectrograph \citep{wilson12,majewski16}
attached to the Sloan 2.5m telescope \citep{gunn06} at Apache Point
Observatory.  A detailed description of the data reduction and
resulting data products can be found in \cite{nidever15} and
\cite{holtzman15}.

We focus on [Fe/H] and abundance ratios whose star to star variations
within GCs are the typical indicators of the presence of multiple
stellar populations, such as [C/Fe], [N/Fe], [Na/Fe], [Al/Fe], and
[Mg/Fe].  While oxygen tends to be low in SG stars, ASPCAP currently
does not provide reliable oxygen abundances for O-poor stars, so
we leave that element out of the analysis in this paper.  The typical
precision of APOGEE abundances is better than 0.1 dex, which is
more than adequate for our present purposes.  The precision of
APOGEE radial-velocities is typically better than $\sim$ 0.5 km/s,
which again exceeds the requirements of our project.

\subsection{Membership}

Once stellar parameters and radial velocities of sample stars
are known, their membership status was further scrutinised by
filtering out all stars failing to meet projected distance,
radial velocity, and metallicity criteria.  In this way, stars with
projected distances from GC centres that are larger than the
GC tidal radii were removed from consideration.  By the same
token, stars with heliocentric radial velocities differing from the
catalogued GC values by more than the GC velocity dispersion (or
$\pm$ 15 km/s in cases where velocity disperison is not available
in the Harris catalog), and those differing in [Fe/H] from the mean
GC values by more than 0.3 dex were also removed from consideration.
The GC centres, tidal radii, and metallicities adopted in these
comparisons were taken from the 2010 version of the \cite{harris96}
catalog.
The only exception is Terzan 5, for which there is a large spread
in [Fe/H], so that no metallicity criterion was adopted.  There is
no evidence for the presence of [Fe/H] variations in the other
sample GCs, so the [Fe/H] criterion adopted was generous enough
that no member stars are expected to be excluded from analysis.
The final list of GC targets is listed in Table~\ref{sample},
together with elemental abundances, radial velocities, and distances
from the host GC centres.

To estimate how many and which stars in the APOGEE sample were
excluded due to adoption of the above selection criteria, we searched
for candidate members meeting relaxed criteria by doubling the
projected distance and radial velocity search ranges.  As a result,
we found that adoption of these more relaxed criteria would have
resulted in addition of a number of stars to our samples for NGC~6553,
6528, and Terzan~5.  In most cases the additional stars fall within
the range of metallicities acceptable for these three GCs, suggesting
possible membership.  However, all the stars in Table~\ref{extra}
have abundance patterns consistent with that of field samples at
the same metallicity, which somewhat reduces the chances that they
are associated with the GCs in our sample.  We nevertheless list
these additional stars in Table~\ref{extra}, for completeness,
although they are not considered in our discussion.

\begin{table*} \label{sample}
 \centering
  \caption{Stellar parameters, elemental abundances, radial velocities, and
cluster-centric distances for sample stars.  Errors reported in all
quantities correspond to pipeline precision estimates.  $T_{\rm eff}$ and
$\log g$ precision in DR12 is $\sim$90~K and $\sim$0.1~dex.}
  \begin{adjustbox}{max width=\textwidth}
  \begin{tabular}{cllcccccccr}
\hline
\hline
APOGEE ID  & $T_{\rm eff}$  &
$\log g$ &  
[Fe/H] & [C/Fe] & [N/Fe] & [Na/Fe] & [Mg/Fe] & [Al/Fe] & {RV (km/s)} & d (\arcmin) \\
\hline
\hline
&&&&&&&&&& \\
&&&&& Palomar 6 &&&&& \\
 2M17434071-2613528 & 3675 & 0.1 &--1.00 $\pm$ 0.04 &--0.02 $\pm$ 0.05  & +0.18  $\pm$ 0.09  & ---             & +0.24 $\pm$ 0.05 &--0.02 $\pm$ 0.09 & 178.621 $\pm$ 0.003 & 0.63 \\
 2M17434331-2610217 & 3983 & 0.8 &--0.85 $\pm$ 0.04 &--0.10 $\pm$ 0.06  & +0.08  $\pm$ 0.09  & ---             & +0.26 $\pm$ 0.05 & +0.14 $\pm$ 0.09 & 175.579 $\pm$ 0.009 & 3.00 \\
 2M17434675-2616068 & 4135 & 1.2 &--0.77 $\pm$ 0.04 &--0.10 $\pm$ 0.06  & +0.69  $\pm$ 0.10  & ---             & +0.14 $\pm$ 0.06 & +0.26 $\pm$ 0.10 & 175.28  $\pm$ 0.01  & 2.94 \\
\hline
&&&&& Terzan 5 &&&&& \\
 2M17475169-2443153 & 3844 & 1.4 &--0.31 $\pm$ 0.03 & +0.15 $\pm$ 0.04  &--0.14  $\pm$ 0.07  & +0.15 $\pm$ 0.12 &--0.22 $\pm$ 0.04 &       ---       &--75.52  $\pm$ 0.02  & 4.60 \\
 2M17480088-2447295 & 3974 & 1.3 &--0.48 $\pm$ 0.03 &--0.34 $\pm$ 0.05  & +1.11  $\pm$ 0.07  & +0.47 $\pm$ 0.11 & +0.04 $\pm$ 0.04 & +0.62 $\pm$ 0.06 &--99.471 $\pm$ 0.006 & 1.17 \\
 2M17480576-2445000 & 3999 & 1.1 &--0.61 $\pm$ 0.04 & +0.02 $\pm$ 0.05  & +0.77  $\pm$ 0.09  &       ---        & +0.21 $\pm$ 0.06 & +0.16 $\pm$ 0.09 &--76.79  $\pm$ 0.02  & 1.76 \\
 2M17480668-2447374 & 3925 & 0.8 &--0.56 $\pm$ 0.04 &--0.39 $\pm$ 0.05  & +1.07  $\pm$ 0.08  & +0.25 $\pm$ 0.12 & +0.12 $\pm$ 0.04 & +0.37 $\pm$ 0.07 &--89.922 $\pm$ 0.008 & 0.96 \\
 2M17481414-2446299 & 3725 & 0.9 &--0.02 $\pm$ 0.03 & +0.09 $\pm$ 0.04  & +0.18  $\pm$ 0.06  & +0.05 $\pm$ 0.09 & +0.04 $\pm$ 0.04 &--0.09 $\pm$ 0.05 &--76.05  $\pm$ 0.01  & 2.12 \\
\hline
&&&&& NGC~6522 &&&&& \\
 2M18032356-3001588 & 4088 & 1.0 &--1.09 $\pm$ 0.04 &--0.38 $\pm$ 0.07  & +1.04  $\pm$ 0.05  &--- & +0.17 $\pm$ 0.06 & +0.40 $\pm$ 0.11 &--13.76 $\pm$ 0.01 & 2.28 \\
\hline
&&&&&&&&&& \\
&&&&& NGC~6528 &&&&& \\
 2M18044775-3003469 & 4167 & 1.7 &--0.16 $\pm$ 0.04 & +0.18 $\pm$ 0.05  & +0.22  $\pm$ 0.07  & +0.23 $\pm$ 0.12 & +0.16 $\pm$ 0.05 & +0.19 $\pm$ 0.07 & 215.40 $\pm$ 0.03 & 0.58 \\
 2M18045107-3002378 & 4168 & 1.7 &--0.21 $\pm$ 0.04 & +0.03 $\pm$ 0.05  & +0.76  $\pm$ 0.08  & +0.61 $\pm$ 0.12 & +0.14 $\pm$ 0.05 & +0.36 $\pm$ 0.07 & 210.75 $\pm$ 0.02 & 0.81 \\
\hline
&&&&& NGC~6553 &&&&& \\
 2M18085726-2558403 & 3762 & 1.3 & +0.20 $\pm$ 0.03 & +0.13 $\pm$ 0.03  &--0.03  $\pm$ 0.05  & ---             & +0.12 $\pm$ 0.03 & +0.11 $\pm$ 0.04 & 9.571  $\pm$ 0.009 & 6.18 \\
 2M18090968-2554574 & 4817 & 2.4 &--0.02 $\pm$ 0.04 &--0.30 $\pm$ 0.06  & +0.74  $\pm$ 0.08  & +0.29 $\pm$ 0.12 &--0.07 $\pm$ 0.05 &       ---       & 5.06   $\pm$ 0.05 & 1.84 \\
 2M18091335-2548357 & 4566 & 2.5 & +0.08 $\pm$ 0.03 & +0.15 $\pm$ 0.05  & +0.25  $\pm$ 0.07  & +0.05 $\pm$ 0.09 & +0.03 $\pm$ 0.04 &--0.06 $\pm$ 0.07 &--8.85  $\pm$ 0.03 & 6.01 \\
 2M18091466-2552275 & 4153 & 1.6 &--0.21 $\pm$ 0.03 &--0.12 $\pm$ 0.04  & +0.97  $\pm$ 0.07  & +0.51 $\pm$ 0.10 & +0.13 $\pm$ 0.04 & +0.41 $\pm$ 0.06 & 2.41   $\pm$ 0.01 & 2.17 \\
 2M18091564-2556008 & 4057 & 1.3 &--0.18 $\pm$ 0.03 &--0.24 $\pm$ 0.04  & +0.91  $\pm$ 0.07  & +0.46 $\pm$ 0.09 & +0.08 $\pm$ 0.04 & +0.47 $\pm$ 0.06 &--1.02  $\pm$ 0.02 & 1.56 \\
 2M18091666-2554424 & 3899 & 1.1 &--0.22 $\pm$ 0.03 &--0.15 $\pm$ 0.04  & +0.79  $\pm$ 0.06  & +0.41 $\pm$ 0.09 & +0.12 $\pm$ 0.04 & +0.31 $\pm$ 0.05 &--11.115 $\pm$ 0.004 & 0.29 \\
 2M18091912-2553326 & 4409 & 1.8 &--0.08 $\pm$ 0.04 &--0.40 $\pm$ 0.05  & +0.98  $\pm$ 0.08  & +0.35 $\pm$ 0.11 & +0.08 $\pm$ 0.05 & +0.36 $\pm$ 0.07 & 3.92   $\pm$ 0.02 & 1.03 \\
 2M18092147-2556039 & 4126 & 1.3 &--0.18 $\pm$ 0.04 &--0.31 $\pm$ 0.04  & +0.96  $\pm$ 0.07  & +0.50 $\pm$ 0.10 & +0.13 $\pm$ 0.04 & +0.39 $\pm$ 0.06 & 7.01   $\pm$ 0.01 & 1.77 \\
 2M18092234-2554381 & 4428 & 2.0 &--0.23 $\pm$ 0.04 & +0.11 $\pm$ 0.05  & +0.26  $\pm$ 0.08  & +0.23 $\pm$ 0.12 & +0.14 $\pm$ 0.05 & +0.06 $\pm$ 0.08 &--0.80  $\pm$ 0.02 & 1.06 \\
 2M18092241-2557595 & 4145 & 1.3 &--0.22 $\pm$ 0.03 & +0.11 $\pm$ 0.05  & +0.19  $\pm$ 0.07  & +0.11 $\pm$ 0.11 & +0.18 $\pm$ 0.05 & +0.41 $\pm$ 0.07 &--7.94  $\pm$ 0.01 & 3.63 \\
 2M18092826-2558233 & 4440 & 2.0 &--0.10 $\pm$ 0.04 &--0.33 $\pm$ 0.05  & +0.90  $\pm$ 0.08  & +0.38 $\pm$ 0.11 & +0.15 $\pm$ 0.05 & +0.39 $\pm$ 0.07 &--1.02  $\pm$ 0.02 & 4.54 \\
 2M18093498-2549038 & 4830 & 2.6 &--0.17 $\pm$ 0.04 & +0.07 $\pm$ 0.06  & +0.06  $\pm$ 0.09  &--0.13 $\pm$ 0.11 & +0.02 $\pm$ 0.05 & +0.28 $\pm$ 0.09 &--15.65 $\pm$ 0.02 & 6.71 \\
\hline
\end{tabular}
\end{adjustbox}
\end{table*}

\begin{table*} \label{extra}
 \centering
  \caption{Stellar parameters, elemental abundances, radial velocities, and
angular cluster-centric distances for stars rejected on the basis of
angular cluster-centric distances and/or radial velocities.  Errors reported in all
quantities correspond to pipeline precision estimates.  $T_{\rm eff}$ and
$\log g$ precision in DR12 is $\sim$90~K and $\sim$0.1~dex.}
  \begin{adjustbox}{max width=\textwidth}
  \begin{tabular}{cllcccccccr}
\hline
\hline
APOGEE ID  & $T_{\rm eff}$  &
$\log g$ &  
[Fe/H] & [C/Fe] & [N/Fe] & [Na/Fe] & [Mg/Fe] & [Al/Fe] & {RV (km/s)} & d (\arcmin) \\
\hline
\hline
&&&&&&&&&& \\
&&&&& Terzan~5 &&&&& \\
2M17480857-2446033 & 3687 & 0.2 &--0.72 $\pm$ 0.04 & +0.00 $\pm$ 0.05 & +0.47 $\pm$ 0.08 &--0.03 $\pm$ 0.15 & +0.24 $\pm$ 0.05 & +0.13 $\pm$ 0.07 &--64.38 $\pm$ 0.01 & 1.09  \\
2M17483971-2452162 & 3938 & 1.0 &--0.21 $\pm$ 0.03 & +0.05 $\pm$ 0.04 & +0.12 $\pm$ 0.07 & +0.00 $\pm$ 0.10 & +0.21 $\pm$ 0.04 & +0.22 $\pm$ 0.06 &--62.50 $\pm$ 0.01 & 9.64 \\
&&&&&&&&&& \\
&&&&& NGC~6528 &&&&& \\
2M18053866-3008454 & 4346 & 1.8 &--0.12 $\pm$ 0.04 &--0.01 $\pm$ 0.05 &--0.07 $\pm$ 0.08 &--0.36 $\pm$ 0.13 & +0.15 $\pm$ 0.05 &--0.36 $\pm$ 0.08 & +185.66 $\pm$ 0.07 & 11.88 \\
&&&&&&&&&& \\
&&&&& NGC~6553 &&&&& \\
2M18082912-2548259 & 3830 & 1.1 & +0.06 $\pm$ 0.03 & +0.01 $\pm$ 0.03 & +0.34 $\pm$ 0.05 & +0.32 $\pm$ 0.07 & +0.00 $\pm$ 0.03 & +0.07 $\pm$ 0.04 & +20.062 $\pm$ 0.002 & 12.50 \\
2M18084319-2547042 & 3678 & 0.4 &--0.37 $\pm$ 0.03 & +0.20 $\pm$ 0.04 & +0.26 $\pm$ 0.07 & +0.24 $\pm$ 0.11 & +0.24 $\pm$ 0.04 & +0.18 $\pm$ 0.06 & +24.817 $\pm$ 0.005 & 10.75 \\
2M18084368-2557107 & 4260 & 1.7 &--0.09 $\pm$ 0.03 & +0.14 $\pm$ 0.05 & +0.15 $\pm$ 0.07 & +0.03 $\pm$ 0.11 & +0.26 $\pm$ 0.05 & +0.21 $\pm$ 0.07 & +3.22   $\pm$ 0.02  & 8.09  \\
2M18085792-2547267 & 3665 & 0.8 &--0.22 $\pm$ 0.03 & +0.27 $\pm$ 0.04 & +0.25 $\pm$ 0.06 & +0.31 $\pm$ 0.11 & +0.20 $\pm$ 0.04 &--0.23 $\pm$ 0.06 &--9.56   $\pm$ 0.01  & 8.35  \\
2M18092650-2541199 & 3885 & 1.3 & +0.07 $\pm$ 0.03 & +0.05 $\pm$ 0.04 & +0.26 $\pm$ 0.05 & +0.22 $\pm$ 0.07 & +0.02 $\pm$ 0.03 &--0.05 $\pm$ 0.04 & +17.440 $\pm$ 0.005 & 13.34 \\
2M18094548-2554255 & 4002 & 1.2 &--0.24 $\pm$ 0.03 & +0.15 $\pm$ 0.04 & +0.21 $\pm$ 0.07 & +0.19 $\pm$ 0.12 & +0.20 $\pm$ 0.04 & +0.19 $\pm$ 0.07 & +19.44  $\pm$ 0.01  & 6.26  \\
2M18095105-2600134 & 3893 & 0.9 &--0.29 $\pm$ 0.03 & +0.16 $\pm$ 0.04 & +0.17 $\pm$ 0.07 & +0.13 $\pm$ 0.12 & +0.30 $\pm$ 0.04 & +0.54 $\pm$ 0.06 &--30.917 $\pm$ 0.009 & 9.43  \\
\hline
\end{tabular}
\end{adjustbox}
\end{table*}

\section{Results}  \label{results}

\begin{figure*}
 \centering
 \includegraphics[width=120mm]{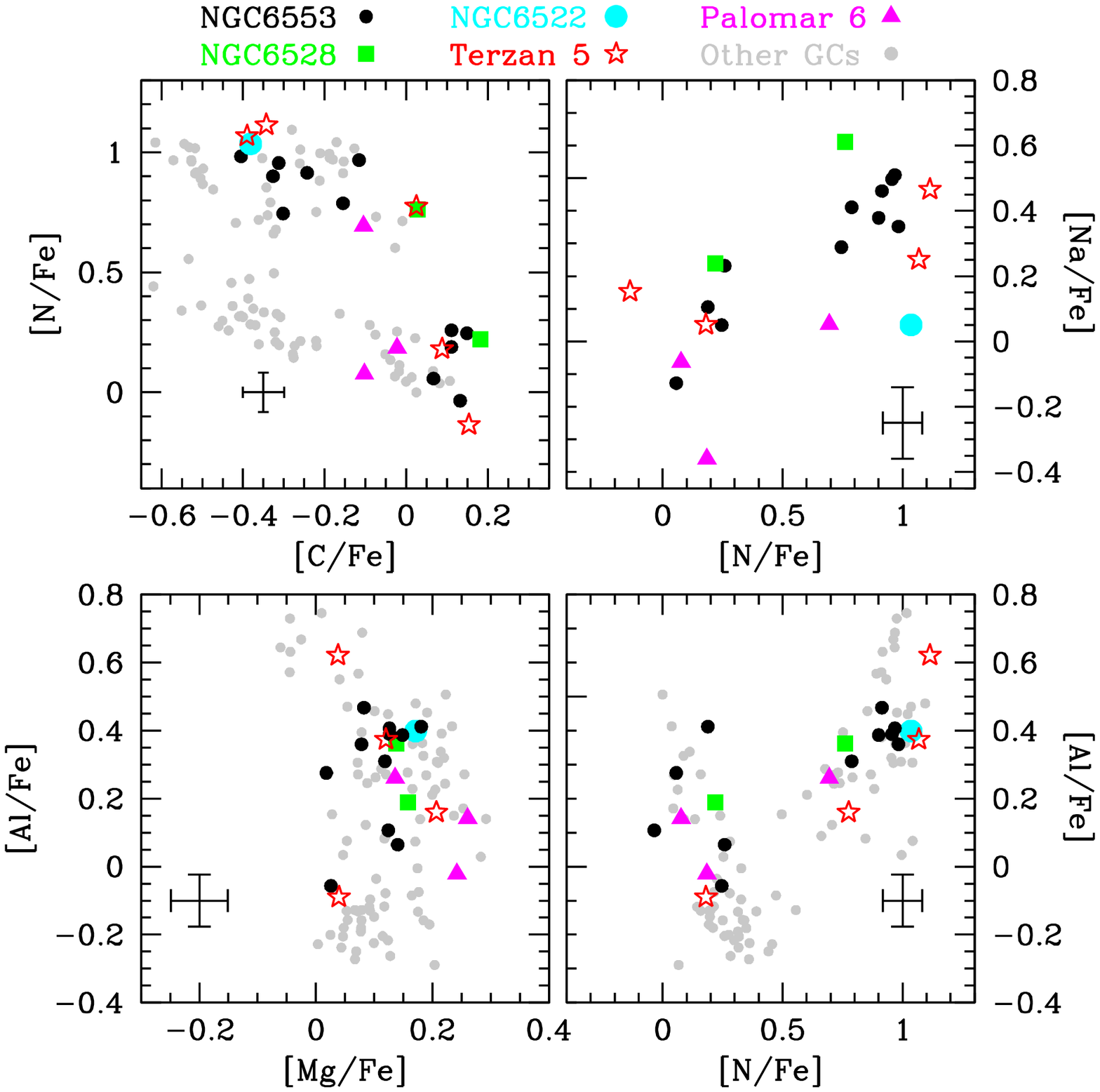}
 \caption{Stars from our GC sample in abundance space.  Symbol types and
 colours are as indicated in the legend.  ``Other GCs'' include DR12 data
 for halo/thick disk M~3, M~5, M~107, M~71, and NGC~6760
\citep{schiavon16}.  See text for details.}
 \label{fig:abs}
\end{figure*}

The key result presented in this article is summarised in
Figure~\ref{fig:abs}, where data for all the members of the program
GCs are displayed in various abundance-ratio planes.  Symbol/colour
codes are adopted to distinguish data for stars from different GCs,
and on each panel the mean error bars are displayed.  Overall, the
stars from different GCs follow consistent trends in each diagram,
exhibiting a clear C-N anti-correlation and also clear Na-N and
Al-N correlations.  As commonly seen in other samples
\citep[e.g.,][]{meszaros15}, there is no clear anti-correlation
between Al and Mg abundances, but rather a substantial spread in
the abundance of the former and a smaller spread in Mg abundances.
A bimodality is clearly seen in [N/Fe] and [C/Fe] but it is not
present in other elemental abundances.  The N-C bimodality is mostly
driven by the data for NGC~6553 for which our sample is largest.
There is a clear spread in the abundances of Al, Na, and Mg, but
no clear sign of a bimodal distribution can be distinguished for
these elements.  To decide whether this difference in behaviour
between different elemental abundances is due to sample size, larger
errors in the abundances of the latter elements, or a real physical
effect, a larger sample will be required.

It is instructive to contrast the data for inner Galaxy GCs with
those for GCs in the outer Galaxy.  The grey circles in
Figure~\ref{fig:abs} indicate the DR12 data for the halo/thick-disk
GCs used by \cite{schiavon16}, which are on average more metal-poor
than the sample discussed in this paper.  The GCs represented by
grey symbols are M~3 ([Fe/H]=--1.5), M~5 ([Fe/H]=--1.3), M~107
([Fe/H]=--1.0), M~71 ([Fe/H]=--0.8), and NGC~6760 ([Fe/H]=--0.4).
Overall, the two sets of GCs occupy the same loci in the various
chemical-composition planes which is reassuring.  But some differences
are noteworthy.  In the C-N plane, there is a large collection of
metal-poor GCs at [C/Fe]$<$--0.2 and [N/Fe]$\simless$+0.5.  This
is due to the presence of a large population of FG stars from M~3
and M~5, which have lower [C/Fe] and slightly higher [N/Fe] than
their counterparts in more metal-rich GCs.  This difference is
likely to be the result of the evolution of C and N abundances in
the Galaxy, a topic that is beyond the scope of this paper.  Regarding
[Al/Fe], there is a hint that the more metal-poor GCs have a larger
spread than their metal-rich counterparts, but that difference is
barely significant given the size of the errors.  The metal-poor
GCs show a possible indication of the presence of a Mg-Al
anti-correlation, as there is a cluster of data with nearly solar
[Mg/Fe] and [Al/Fe]$\sim$+0.6.  Those stars all belong to M~3, so
they are fairly metal-poor.  Interestingly, one of the Terzan~5
stars in our sample inhabits the same region of the Mg-Al plane.
With [Fe/H]=--0.48, this star is ten times more metal-rich than M~3
members, suggesting that the same process leading to the Mg-Al
anti-correlation may be present at high metallicity.  Finally, we
note that data for [Na/Fe] are not shown for the metal-poor GCs,
since that elemental abundance is quite uncertain in APOGEE DR12
for metal-poor GCs.

\begin{figure}
 \centering
 \includegraphics[width=90mm]{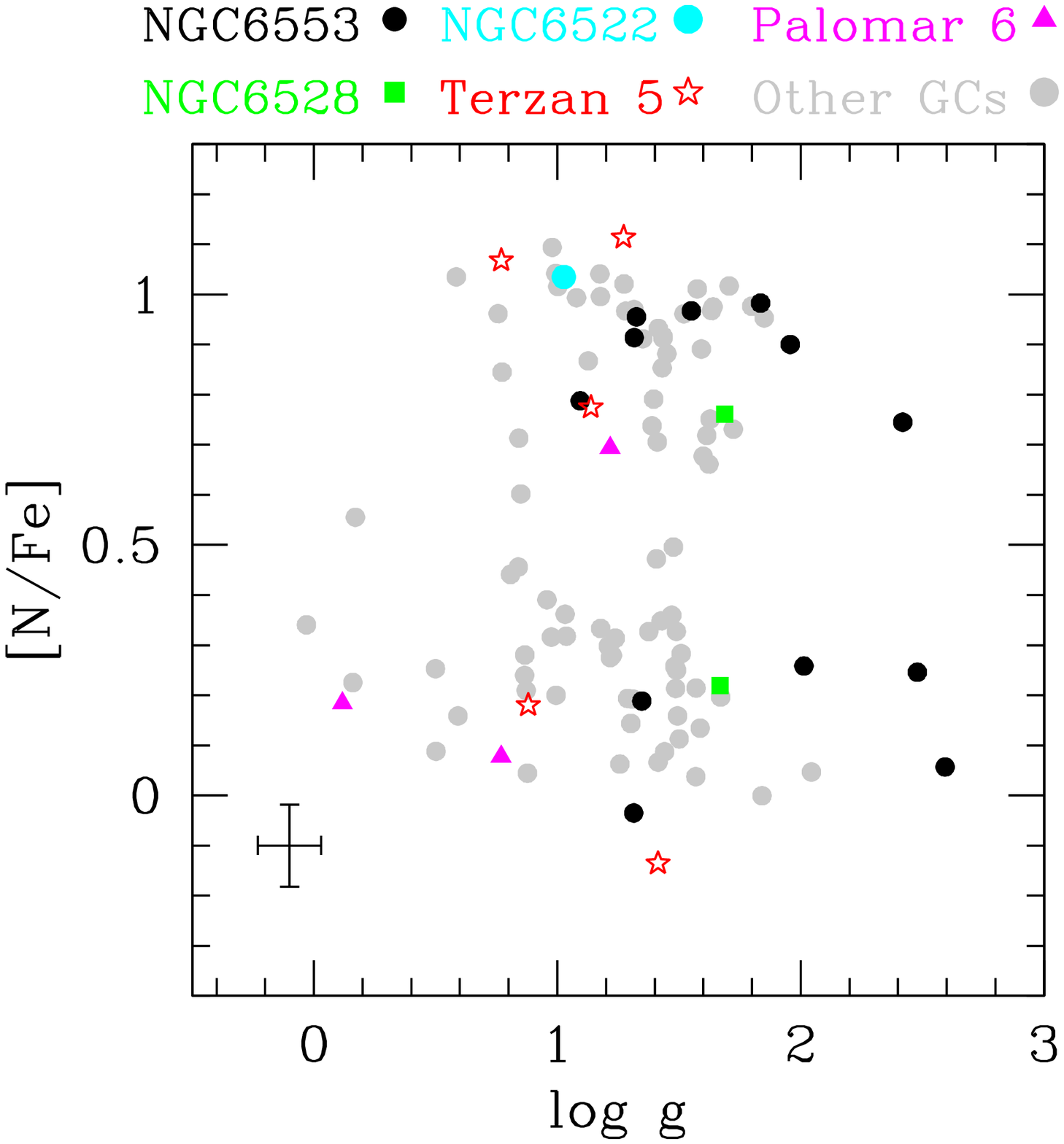}
 \caption{Nitrogen abundance ratios as a function of surface gravity for
the sample stars.  Stars of approximately the same gravity (thus same
evolutionary stage) have vastly different [N/Fe], which cannot be explained
by mixing.}
 \label{fig:Nlog}
\end{figure}

The abundances of some elements, such as nitrogen and carbon, are
known to vary during evolution along the giant branch, both due to
the first dredge-up and extra mixing further up that evolutionary
sequence.  It is important to distinguish the star-to-star abundance
variations reported in Figure~\ref{fig:abs} from those due to stellar
evolution effects.  Figure~\ref{fig:Nlog} shows our sample in the
[N/Fe]--$\log g$ plane.  Stars from all GCs exhibit strong [N/Fe]
differences at near constant $\log g$, which shows that stars at
nearly the same evolutionary stage have vastly different [N/Fe]
abundances, which argues against evolutionary effects.  We conclude that
the intra-GC abundance variations reported in Figure~\ref{fig:abs} are
indicative of the presence of multiple populations in our sample GCs.

It is the first time that the presence of spreads in the abundances
of light elements has been established for the GCs in our sample,
although for at least one of them (NGC~6528) it has been previously
suggested by studies based on medium-resolution spectroscopy
\citep{martell09}.  Such data provide valuable constraints on model
predictions in a regime where they have not been sufficiently tested.
In particular, the presence of large star-to-star abundance
variations---especially in the case of Al---for high-metallicity
GCs challenge model predictions suggesting that the amplitude
of star-to-star variations should decrease towards high metallicity
\citep[e.g.,][]{karakas10,ventura13,bastian15,dicriscienzo16}.  A
detailed confrontation between our data and model predictions is
beyond the scope of this paper.  In the remainder of this Section
we briefly discuss the data for each GC separately.

\subsection{NGC 6553}

One of the most metal-rich GCs known in the Galaxy with [Fe/H]$\sim$--0.2
\citep[e.g.,][]{cohen99,origlia02,alvesbrito06,johnson14,tang16},
NGC~6553 is moderately massive
\citep[$3\times10^5M_\odot$][]{mclaughlin05}, so one would expect
it to host a measurable spread in the abundances of Na \citep{carretta10}
and N \citep{schiavon13}.  It is also the GC for which our sample
is the largest, so the finding of a large spread in the abundances
of N ($\sim$ 1 dex), Na ($\sim$ 0.5 dex), C ($\sim$ 0.5 dex), and
Al ($\sim$ 0.5 dex) is statistically robust.  Interestingly, Al
abundances in a couple of the N-normal stars are as high as those
in their N-rich counterparts.  Except for the presence of a few
outliers that are probable non-members \citep[see][]{tang16}, no
clear spread in the abundances of heavy elements such as Fe or Ca
is detected.  The abundances of N, C, and perhaps Na seem to be
bimodal, but any firm conclusion should await a substantial increase
in sample size.  In a separate publication \citep{tang16}, a careful
evaluation of membership and a quantitative comparison between model
predictions and observations is pursued for the case of NGC~6553,
for which our stellar sample is the largest.  We refer the reader
to that paper for a more detailed analysis of the APOGEE data for
that cluster.

\subsection{Terzan 5}

Terzan~5 is a peculiar object.  Its colour-magnitude diagram displays
two well-separated red horizontal branches \citep{ferraro09}, and
it is known to host a very large population of millisecond pulsars
\citep{ransom05}.  It is one of the few Galactic GCs for which a
considerable spread in [Fe/H] and possibly age has been detected
\citep[e.g.,][]{ferraro09,ferraro16}.  Indeed, \cite{origlia13} and
\cite{massari14a,massari14b} used Keck/NIRSPEC, Keck/DEIMOS, and
VLT/FLAMES spectra to establish the presence of at least three
well defined stellar populations in Terzan~5, with [Fe/H]$\sim$--0.8, --0.3, and
+0.2 (the highest metallicity detected in any Galactic GC.
Interestingly, \cite{origlia11} found that the relation between the
abundances of Fe and the $\alpha$ elements Mg and O mimics that of the
Galactic bulge itself.  Unlike other massive Galactic GCs that show
both a spread in Fe and light-element abundances \citep[for a review
see][]{dacosta15}, no spreads and anti-correlations between the abundances
of light elements were previously detected in Terzan~5
\citep[e.g.,][]{origlia11,origlia13}.  This is surprising given its
high mass \citep[$\sim2\times10^6M_\odot$;][]{ferraro12} and the
high metallicity of some of its stellar populations.  These results led to
the suggestion that Terzan~5 is the remnant core of a dwarf galaxy,
or perhaps even an early fragment of the original bulge formation
\citep{ferraro12}.

Our sample for Terzan~5 consists of only five stars, yet the data
reveal a large spread in light-element abundances, including a
strong N-C anti-correlation and equally strong N-Na and N-Al
correlations.  Membership uncertainties are obviously a concern in
this case.  All five stars fall safely within the metallicity range
established by \cite{massari14b}, although the N-rich sub-sample
is particularly metal-poor, especially considering a zero-point
correction of --0.2~dex to APOGEE raw [Fe/H] \citep{holtzman15}.
\cite{origlia13} measured the systemic radial velocity of Terzan~5
to be --82~km/s, with a velocity dispersion of $\sigma\sim$15~km/s.
All but one of our sample stars are situated within 1$\sigma$ of
Terzan~5's systemic radial velocity.  The exception is the N-rich
star {\tt 2M17480088-2447295}, with $v_r=-99.5$km/s---which is too
low by $\sim$2.5km/s to be within 1$\sigma$ from the GC systemic RV.
Since this star has a very high [N/Fe] and, at a distance of
$\sim$1\arcmin\ from the cluster centre, is well within its tidal
radius ($\sim$6.\arcmin7), we deem it a very likely member of
Terzan~5.  

ASPCAP results for some of the Terzan~5 candidate members merit
a more detailed look.  Some of the elemental abundances for star
{\tt 2M17475169-2443153} are quite unusual, with [N/Fe]=--0.14,
[Mg/Fe]=--0.22 and [Al/Fe]=--1.  Visual inspection showed that there
is an important mismatch between synthetic and observed spectra at
the continuum level through a large fraction of the APOGEE spectral
interval.  In those regions, the normalized flux in continuum pixels
of the observed spectrum is higher than unity and higher than their
counterparts in the synthetic spectrum.  This mismatch is probably
due to a problem in the normalization of the observed spectrum.
Such a mismatch is most likely responsible for an underestimate of
the N, Na, and Mg abundances, as lines due to CN (e.g.,
$\lambda\lambda$15322, 15332${\rm\AA}$) and Mg (e.g.,
$\lambda\lambda$15339, 15959${\rm\AA}$) are affected by this issue
particularly strongly.  Interestingly, an acceptable match of the
observed spectrum by ASPCAP is found in regions of known OH (e.g.,
$\lambda\lambda$16708, 16719, 16889${\rm\AA}$) and CO (e.g.,
$\lambda\lambda$15368, 15997, 16189${\rm\AA}$) lines \citep{garcia16}.
Perhaps most importantly, such a systematic effect on the continuum
normalization could also potentially lead to an overestimate of
surface gravity, which may impact the values of abundances inferred
from molecular lines which are sensitive to $\log g$.  Indeed, the
ASPCAP surface gravity for this star ($\log g = 1.4$) is quite high
for such a cool star ($T_{\rm eff}$=3844K), which could be the
result of the continuum systematics identified above.  Without
further calculations it is impossible to gauge the impact of this
effect on the chemical composition of this star.  Given the good fit to CO
and OH lines, the abundance of C is probably reliable, modulo $\log g$
effects.  On the other hand, we consider the abundances of N, Mg, and Na 
to be questionable.

Regarding Al, the situation is more complex, as ASPCAP reports
[Al/Fe]=--1 for {\tt 2M17475169-2443153}, which we find suspiciously
low.  The strengths of all three Al lines in the APOGEE spectrum
($\lambda\lambda$16723, 16755, and 16767${\rm\AA}$) are overpredicted
by ASPCAP, suggesting that [Al/Fe] for this star should be even
lower.  However, comparison of the spectrum of {\tt 2M17475169-2443153}
with those of stars with very similar stellar parameters and CNO
abundances (e.g., {\tt 2M06182536+3414581, 2M17493226-2309585,
2M18322950-1246417}, and {\tt 2M18493324-0302028}) but vastly
different [Al/Fe] (ranging from --0.1 to -0.59) showed very similar
line strengths, which is surprising.  We do not understand the
origin of these issues, so we choose not to consider the abundances
of Mg, Na and Al for this star.

We also examined the abundances of star {\tt 2M17480576-2445000},
particularly due to the fact that ASPCAP found a very low value for
[Na/Fe]=--0.47, which is surprising due to its very large [N/Fe]=+0.77.
Inspection of the two Na lines in the APOGEE region suggests that
[Na/Fe] for this star is not reliable.  The strongest of the two
lines ($\lambda$16393${\rm\AA}$) is lost to bad pixels, whereas the
remaining weaker line ($\lambda$16378${\rm\AA}$) is severly
underestimated by ASPCAP, despite the fact that the observed continuum
is well matched by the best fitting synthetic spectrum.  On the
other hand, ASPCAP does an excellent job of matching the CN lines
in the spectrum of this star, so that we consider that [N/Fe] is
quite reliable.  We therefore decide to ignore the abundance of Na
for this star in the remainder of our analysis.


We conclude that there is strong evidence for the presence of an
intrinsic spread, as well as correlations and anti-correlations
between light-element abundances in Terzan~5.  A more detailed
study, based on a larger sample and including additional elemental
abundances is in order.

\subsection{NGC 6528}

With [Fe/H]$\sim$--0.2, NGC~6528 is another very metal-rich bulge
GC \citep{carretta01,barbuy04,zoccali04,origlia05}.  It is also
moderately massive \citep[$\sim 2\times 10^5 M_\odot$;][]{mclaughlin05}.
Based on medium-resolution spectroscopy of a relatively small sample,
\cite{martell09} determined the presence of a bimodal distribution
of CN band strengths.  In contrast, \cite{calamida14} obtained
Str\"omgren photometry for a larger sample, but did not report the
presence of a bimodal distribution in CN-sensitive colour indices.
Our sample contains only two NGC~6528 candidate members, with 
essentially identical atmospheric parameters, but a significantly
different N and C, and Na abundances suggesting the presence of an
abundance spread in this GC as well, in agreement with the result
by \cite{martell09}.

\subsection{Palomar 6}

\cite{lee04} determined the metallicity of Palomar 6 to be
[Fe/H]$\sim$--1.  The cluster is moderately massive \citep[$\sim2.3\times
10^5 M_\odot$;][]{boyles11}, and therefore we would expect in this case
to detect the presence of multiple populations.  The sample of
candidate members is small, and the only element found to present
variations that are statistically significant is N.  While
there are indications that C and Al also show variations, they are
not significantly larger than the error bars.  The data on Na, on
the other hand, suggest the presence of variations.  However, at
the relatively low metallicity of Palomar 6, Na abundances in DR12
are uncertain, due to the weakness of the available lines in the
APOGEE spectrum.  We conclude that a definite variation in [N/Fe]
was detected, but for C, Al, and Na more data are required to
establish the definitive presence of abundance spreads.

\subsection{NGC 6522}

An old GC located in the inner Galaxy, NGC~6522 is moderately metal
poor with [Fe/H]$\sim$--1 \citep{barbuy09} and has a low mass
\citep[$\sim 6\times10^4M_\odot$,][]{gnedin97}.  It is thus not
immediately clear that NGC~6522 is expected to contain multiple
populations.  Our sample contains only one candidate member of this
GC, which makes any conclusion on the presence of multiple populations
by definition very uncertain.  However, the star is likely a member
on the grounds of metallicity, radial velocity, and position, and
it has a very high nitrogen abundance ([N/Fe]=+1.04).  Moreover,
in Figure~\ref{fig:Nlog} one can see that our NGC~6522 sample star
occupies the same locus in [N/Fe]--$\log g$ space as second generation
stars in other GCs.  Indeed, at the metallicity of NGC~6522, only
SG stars attain such high nitrogen abundances in the APOGEE DR12
data \citep[][Figure 1]{schiavon16}.  Therefore, we conclude that
NGC~6522 is likely to host multiple populations, but a strong
statement to that effect must await the availability of good quality
data for a larger sample.

%
%

\section{Conclusions}  \label{conclusions}

As part of the SDSS-III/APOGEE survey, we have obtained elemental
abundances for 23 candidate members of five massive globular clusters
located in the inner Galaxy, including some of the most metal-rich
known Galactic GCs.  Our main conclusions are summarised as follows:

\begin{itemize}

\item Spreads in [N/Fe] have been detected in all GCs for which
more than one star was observed, and all these abundances could be
measured.  Among the latter GCs, all but Palomar 6 also exhibit
significant spreads in [C/Fe], [Al/Fe], and [Na/Fe].  The standard
anti-correlation between N and C, and correlations between N and
Al and Na are also present in our data.  This result indicates the
prevalence of the multiple-population phenomenon in GCs as metal-rich
as [Fe/H]$\sim$--0.1.  In at least one GC for which our sample is
largest (NGC~6553), there is strong evidence for a bimodal
distribution in the abundances of C and N.  We conclude that N-rich
stars are present in metal-rich GCs, and therefore they can be used
as reliable tracers of field GC-like populations over a wide range
of metallicities.  As a corollary, the absence of a large
population of metal-rich N-rich stars in the sample discovered by
\cite{schiavon16} in the inner Galaxy reflects the metallicity
distribution function of the presumptive destroyed GCs;

\item The presence of large spreads in [N/Fe] in metal-rich GCs is
interesting also from the point of view of stellar-evolution models,
as predictions from different groups seem to imply a strong dependence
on metallicity of the yields of certain elements
\citep[e.g.,][]{karakas10,ventura13,dicriscienzo16}.  A large spread
of Al abundances in GCs of near-solar metallicity may constitute
an important challenge to existing models.  It would be naive to
rule out a metallicity dependence of stellar yields on the basis
of our results alone, but hopefully our numbers can be used for
careful comparison with detailed model predictions based on
state-of-the-art yields;

\item For most of the GCs in our study the presence of multiple
populations has been uncovered for the first time.  This is the
case for NGC~6553 \citep[see also][]{tang16}, NGC~6528, and Palomar~6.
In the case of NGC~6522, the sample contains only one star, which
happens to have a SG abundance pattern.  Since no GC is known to
host only SG stars, it is very likely that NGC~6522 also contains
multiple populations.  For Terzan~5, although multiple populations
were known to be present with different [Fe/H], it is the first
time that star-to-star variations in light elements is reported;


\item The case of Terzan~5 is particularly interesting, as it is a
very massive GC with a substantial spread in [Fe/H]---a feature
that places it in the same category as objects like $\omega$~Cen, M54,
and others, which are often referred to as remnants of dwarf
satellites accreted to the Galaxy \citep[for a short review,
see][]{dacosta15}.  It is interesting, in this regard, that Terzan~5
displays abundance spreads and correlations similar to those found in
$\omega$~Cen \citep{johnson10}, which suggests that these systems
may occupy a region of parameter space somewhere in-between
globular clusters and dwarf spheroidal galaxies.

\end{itemize}

The pursuit of a solution to the enigma of globular cluster formation
and its connections with the formation of the Galaxy itself will
receive a great boost once a massive database of chemistry and
kinematics of members of {\it all} Galactic GCs is in place.  Despite
valiant efforts, we are still scratching the surface.  Further
progress is expected from observations taken from the Southern
hemisphere with APOGEE-2 \citep[e.g.,][]{majewski16}.  In addition,
the advent of high-resolution NIR spectrographs with large multiplexing
power and small fiber-collision radii, such as MOONS \citep{cirasuolo14},
will potentially bring about a paradigm shift in this field.

\section*{Acknowledgments}

R.P.S. thanks Maurizio Salaris and Nate Bastian for useful comments
on an early version of the manuscript.  The anonymous referee is
thanked for a careful reading of the manuscript and valuable
suggestions.  R.E.C. acknowledges support from Gemini-CONICYT for
Project 32140007".  S.L.M. acknowledges support from Australian
Research Council grant DE140100598.  Sz.M. has been supported by
the Premium Postdoctoral Research Program of the Hungarian Academy
of Sciences, and by the Hungarian NKFI Grants K-119517 of the
Hungarian National Research, Development and Innovation Office.
D.G., and B.T. gratefully acknowledge support from the Chilean BASAL
Centro de Excelencia en Astrof\'\i sica y Tecnolog\'\i as Afines
(CATA) grant PFB-06/2007.  S.R.M. thanks NSF grant AST-1312863
T.C.B. acknowledges partial support for this work from grants PHY
08-22648; Physics Frontier Center/Joint Institute for Nuclear
Astrophysics (JINA), and PHY-1430152; Physics Frontier Center/JINA
Center for the Evolution of the Elements (JINA-CEE), awarded by the
US National Science Foundation.  D.A.G.H. was funded by the Ram\'on
y Cajal fellowship number RYC-2013-14182 and he acknowledges support
provided by the Spanish Ministry of Economy and Competitiveness
(MINECO) under grant AYA-2014-58082-P.

Funding for SDSS-III has been provided by the Alfred P. Sloan
Foundation, the Participating Institutions, the National Science
Foundation, and the U.S. Department of Energy Office of Science.
The SDSS-III web site is {\tt http://www.sdss3.org/}.  SDSS-III is
managed by the Astrophysical Research Consortium for the Participating
Institutions of the SDSS-III Collaboration including the University
of Arizona, the Brazilian Participation Group, Brookhaven National
Laboratory, University of Cambridge, Carnegie Mellon University,
University of Florida, the French Participation Group, the German
Participation Group, Harvard University, the Instituto de Astrof\'\i
sica de Canarias, the Michigan State/Notre Dame/JINA Participation
Group, Johns Hopkins University, Lawrence Berkeley National Laboratory,
Max Planck Institute for Astrophysics, New Mexico State University,
New York University, Ohio State University, Pennsylvania State
University, University of Portsmouth, Princeton University, the
Spanish Participation Group, University of Tokyo, University of
Utah, Vanderbilt University, University of Virginia, University of
Washington, and Yale University.

\label{lastpage}

\end{document}